\documentclass[ aps, showpacs, showkeys, nofootinbib, floatfix, superscriptaddress]{revtex4}

\usepackage{amsfonts}

\usepackage{amssymb}
\usepackage{amsmath}
\usepackage{graphicx}

\begin{document}

\title{Reduced modified Chaplygin gas cosmology}

 \author{Jianbo Lu}
 \email{lvjianbo819@163.com}
 \affiliation{Department of Physics, Liaoning Normal University, Dalian 116029, P. R. China}
  \author{Danhua Geng}
 \affiliation{Department of Physics, Liaoning Normal University, Dalian 116029, P. R. China}
 \author{Lixin Xu}
\affiliation{School of Physics and Optoelectronic Technology, Dalian University of Technology, Dalian, 116024, P. R. China}
 \author{Yabo Wu}
 \affiliation{Department of Physics, Liaoning Normal University, Dalian 116029, P. R. China}
 \author{Molin Liu}
 \affiliation{College of Physics and Electronic Engineering, Xinyang Normal University, Xinyang 464000, PR China}

\begin{abstract}
In this paper, we study cosmologies containing the reduced modified Chaplygin gas (RMCG) fluid which is reduced from the modified Chaplygin gas $p=A\rho-B\rho^{-\alpha}$ for the value of $\alpha=-1/2$. In this special case, dark cosmological models can be realized for different values of model parameter $A$. We investigate the viabilities of these dark cosmological models by discussing  the evolutions of cosmological quantities  and using the currently available cosmic observations.  It is shown that the special RMCG model ($A=0$ or $A=1$) which unifies the dark matter and dark energy should be abandoned. For $A=1/3$, RMCG which unifies the dark energy and dark radiation is the favorite model according to the objective Akaike information criteria. In the case of $A<0$, RMCG can achieve the features of the dynamical quintessence and phantom models, where the evolution of the universe is not sensitive to the variation of model parameters.

\end{abstract}



\maketitle

\section{$\text{Introduction}$}

{Observations indicate some challenges to the standard Big Bang model of cosmology. Several invisible components what we have to search  in universe are hinted. For example, the observations on rotation curve of galaxy \cite{DM-rotation}  directly  relate  to the amount of pressureless matter, proposing   dark matter (DM)  in our Universe; The observations on  supernovae of type Ia \cite{SNIa,SNIa1} point out an accelerating universe at late time, which is usually interpreted as the existence of a new ingredient  called dark energy (DE);  The Wilkinson microwave anisotropy probe (WMAP) provides  precise measurement of the cosmic microwave background radiation. Combining the  9-year WMAP results with the Hubble constant measured from the Hubble Space Telescope (HST) and the baryons acoustic oscillations (BAO) from the SDSS puts a constraint on the effective number of relativistic degrees of freedom $N_{eff}=3.84\pm 0.4$ which implies the presence of an extra dark radiation (DR) component at 95\% confidence level  \cite{9ywmap,9ywmap1}. \footnote{Recently, Ref. \cite{DR-Hprior} studied the effect of $H_{0}$ prior on the value of  $N_{eff}$. In the $\Lambda$CDM model, the evidence of DR is weakened to $\sim 1.2$ standard deviations ($N_{eff}=3.52 \pm 0.39$ at 68\% confidence level) \cite{DR-Hprior}  by taking the median statistics (MS) prior  $H_{0}=68 \pm 2.8$ km s$^{-1}$Mpc$^{-1}$ to replace the HST  prior  $H_{0}=73.8 \pm 2.4 $ km s$^{-1}$Mpc$^{-1}$. This result tends to show that the evidence for DR is not pressing any more. }. It is interesting  to search  origins of these dark sectors. In the past years, efforts were made to study these dark sectors comprising the DM, DE and DR, such as the seeking for the candidates of the cold and warm dark matter \cite{DM,DM1}, the discussion on the  cosmological constant and the dynamical DE \cite{DEmodels0,DEmodels1,DEmodels2,DEmodels3,DEmodels4,DEmodels5,DEmodels6,DEmodels7,DEmodels8,DEmodels9,DEmodels10,DEmodels11,DEmodels12}, the exploration for the origins of DR using the decayed particle \cite{DR-decaying,DR-netrino}, the interacting DM \cite{DR-interaction}, the Horava-Lifshitz gravity \cite{DR-HL,{DR-HL}1} and extra dimensions \cite{DR-brane}, {\it etc.}.

  In addition to these dark sectors (DM, DE, DR), baryon and radiation as visible constituents naturally exist in our Universe. Current cosmic observations suggest that our Universe contains about 70\% the negative-pressure DE, 30\% the pressureless matter (or called dust) including the DM and baryon, and a small fraction of radiation components which are composed of the photon, neutrino  as well as additional  relativistic species \cite{DR-13035076}.   Someone proposed an economical model which can unify the DM and DE in  a single fluid, say the generalized Chaplygin gas \cite{GCG,GCG1,GCG2,GCG-perturbation} and the modified Chaplygin gas (MCG) \cite{MCG,MCG-problem} for instances. In this paper we will perform  new search of dark sectors from the reduced MCG  (RMCG) fluid. We study the RMCG fluid using the analyses of theoretical constraints and the comparisons with the observational data, and obtain several interesting properties such as the DE and DR can be uniformly described by this single fluid, the evolutions of  the cosmological quantities in the dynamical RMCG model are not sensitive to the variation of model-parameter values, and so on.

This paper is organized as follows. In the next section, we introduce the dark models in the RMCG cosmology.   In Sec. III, we examine the evolutions of growth factor and Hubble parameter in the RMCG model, and compare them with the current observational data. The parameter evaluation and model comparison for the RMCG model are performed in Sec. IV. Sec. V is the conclusions.

\section{$\text{Dark models in RMCG cosmology}$}

The MCG model was widely studied for explaining the cosmic inflation \cite{mcg-inflation,mcg-inflation1,mcg-inflation2,mcg-inflation3} or providing an unified model of the DM and DE \cite{mcg-um,mcg-um1,mcg-um2,mcg-um3}. We consider the equation of state (EoS)
\begin{equation}
p=A\rho-B\rho^{1/2},\label{p-EU}
\end{equation}
dubbed as the RMCG, which is reduced from the modified Chaplygin gas $p=A\rho-B\rho^{-\alpha}$ for the constant model parameter $\alpha=-1/2$. This model (\ref{p-EU}) can produce a  emergent universe  without the time singularity \cite{rmcg-EU,rmcg-EU1,rmcg-EU2,rmcg-EU3}. But in this paper, we will take this RMCG fluid as the dark components in our Universe.

Using the energy conservation equation $d\rho/dt=-3H(\rho+p)$, we obtain the energy density of the RMCG fluid,
\begin{eqnarray}
\rho_{RMCG}(a)&=& [\frac{B}{(1+A)}+\frac{C}{1+A}a^{\frac{-3(1+A)}{2}}]^{2}\nonumber\\
&&=\rho_{0RMCG}[A_{s}^{2}+(1-A_{s})^{2}a^{-3(1+A)}+2A_{s}(1-A_{s})a^{\frac{-3(1+A)}{2}}]\nonumber\\
&&=\rho_{1}+\rho_{2}a^{-3(1+A)}+\rho_{3}a^{\frac{-3(1+A)}{2}},\label{density-EU}
\end{eqnarray}
where $C$ is an integration constant, $A_{s}=B\rho_{0RMCG}^{-1/2}/(1+A)$. $\rho_{1}$, $\rho_{2}$ and $\rho_{3}$ are current values of three energy densities in the RMCG fluid. According to Eq. (\ref{density-EU}), some unified models can be achieved for different values of parameter $A$. Fixing $A$ to zero, we have a  unified model containing the DM, DE and cosmic component having  $w=p/\rho=-1/2$. For $A=1$, the RMCG unifies the DE, DM and stiff matter ($w=1$). In the case of $A=1/3$, a  unified  model including the DE, DR and exotic component ($w=-1/3$) can be arrived. If $A$ is a free positive model parameter ($A\neq 0, 1,1/3$), we obtain a unified model comprising the DE and an unknown  component. In the range of $A<0$, RMCG fluid plays the role as the phantom-like ($A<-1$) and quintessence-like ($0>A>-1$) dynamical DE. In a spatial flat Friedmann-Robertson-Walker (FRW) universe containing  the RMCG fluid, one has the
Friedmann equation
\begin{eqnarray}
H^{2}(a)/H_{0}^{2}&=&\Omega_{0i}a^{-3(1+w_{i})}+\Omega_{RMCG}(a) \nonumber\\
&&=\Omega_{0i}a^{-3(1+w_{i})}+(1-\Omega_{0i})[A_{s}^{2}+(1-A_{s})^{2}a^{-3(1+A)}+2A_{s}(1-A_{s})a^{\frac{-3(1+A)}{2}}]\nonumber\\
&&=\Omega_{0i}a^{-3(1+w_{i})}+\Omega_{01}+\Omega_{02}a^{-3(1+A)}+\Omega_{03}a^{\frac{-3(1+A)}{2}},\label{H2}
\end{eqnarray}
where $\Omega_{0i}$ is the current dimensionless energy density beyond the dark sectors, $\Omega_{01}$, $\Omega_{02}$ and $\Omega_{03}$ correspond to three current dimensionless energy densities given by the RMCG fluid. $a$ is the scale factor that is related to cosmic redshift by $a=1/(1+z)$. In the following, we  show  expressions of  some basic cosmological parameters in the RMCG model:

(1) The adiabatic sound speed  for the RMCG fluid, $c_{s}^{2}=\delta p/\delta \rho=A-\frac{\frac{1}{2}(1+A)A_{s}}{A_{s}+(1-A_{s})a^{-\frac{3}{2}(1+A)}}$. A small non-negative sound speed for matter component  is necessary for forming the large scale structure of our Universe.

(2)  Equation of state for the RMCG fluid, $w=p/\rho=A-\frac{(1+A)A_{s}}{A_{s}+(1-A_{s})a^{-\frac{3}{2}(1+A)}}$. To obtain  a late time accelerating expansion universe, it should be respected that the current value of EoS $w_{0}<-\frac{1}{3}$.   Table \ref{theoretical-constraint} lists the theoretical constraints on model parameter $A_{s}$ in the RMCG cosmology by locating the $w_{0}$ at the quintessence region or phantom region, where the different values or intervals for model parameter $A$ are adopted.

(3) Deceleration parameter $q(a)=-\ddot{a}/(aH^{2})$. An  expanding universe having a transition from deceleration to acceleration is consistent with the current cosmic observations.

(4) Dimensionless density parameter $\Omega_{j}=\rho_{j}/\rho_{c}$. $\rho_{c}=3H^{2}/(8\pi G)$ is the critical density, and $j$ denotes the energy component in our Universe.

  \begin{table}[!htbp]
 \vspace*{-12pt}
 \begin{center}
 \begin{tabular}{ | c | c |c |c | c | c |} \hline\hline
                                & $A=1$    &   $A=\frac{1}{3}$  &  $A=0$   & $-1<A<0$   &   $A<-1$ \\\hline
   $-1< w_{0}<-\frac{1}{3}$     & $1> A_{s}>\frac{2}{3}$          &  $1> A_{s}>\frac{1}{2}$      &  $1> A_{s}>\frac{1}{3}$
                                & $1> A_{s}>\frac{1+3A}{3(1+A)}$  & $\frac{1+3A}{3(1+A)}>A_{s}> 1$     \\\hline
   $w_{0}<-1$                   & $A_{s}>1$    & $A_{s}>1$      &  $A_{s}>1$     & $A_{s}>1$   &   $A_{s}<1$   \\\hline\hline
 \end{tabular}
 \end{center}
 \caption{ Theoretical constraints on RMCG model parameter $A_{s}$ by assuming   $-1< w_{0}<-1/3$ (quintessence)  and $w_{0}<-1$ (phantom), where the different values or intervals for parameter $A$ are adopted in prior. }\label{theoretical-constraint}
 \end{table}

\subsection{$\text{ Should the unified model of DE and DM   be ruled out in RMCG cosmology}$}

\begin{figure}[ht]
  \includegraphics[width=4.3cm]{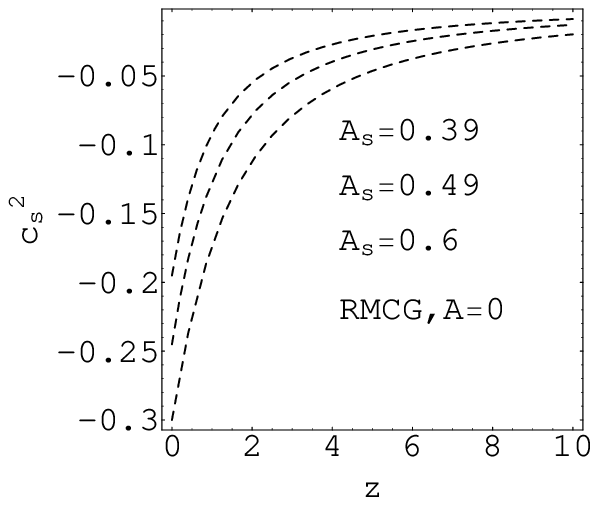}
  \includegraphics[width=4.3cm]{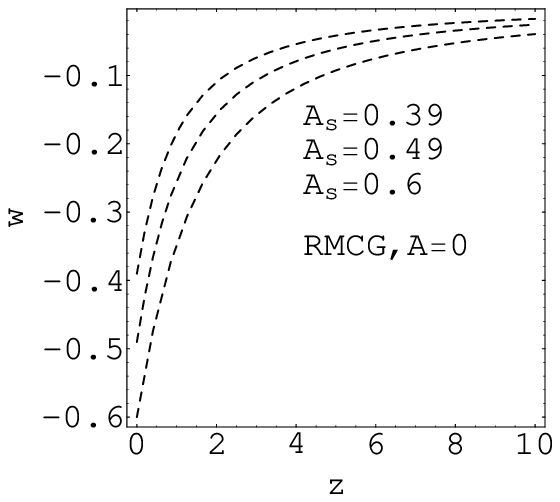}
  \includegraphics[width=4.3cm]{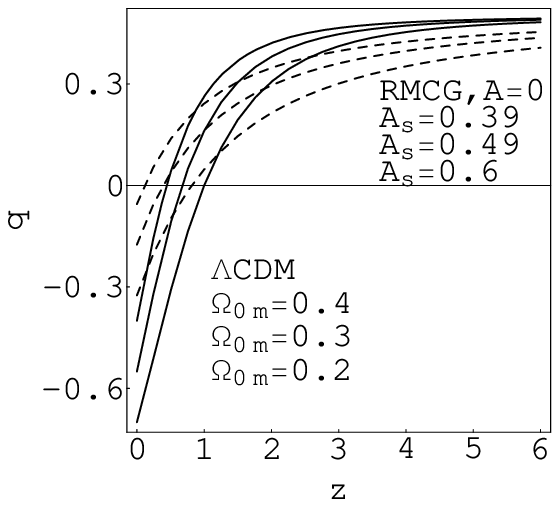}
   \includegraphics[width=4.3cm]{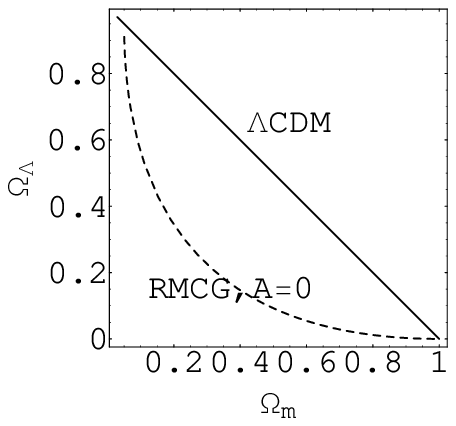}\\
  \caption{Evolutions of the adiabatic sound speed $c_{s}^{2}(z)$, EoS $w(z)$ and  deceleration parameter $q(z)$, and values of dimensionless density parameters for the RMCG ($A=0$) model. Solid lines depict the case of the $\Lambda$CDM. }\label{cs2-omega-A=0}
\end{figure}

For $A=0$ or $A=1$,  a  unified model of DE and DM can be obtained. In the case of $A=0$, the RMCG fluid includes the DM, DE and new hinted dark ingredient $(w=-1/2)$, where the Friedmann equation is written as
 \begin{eqnarray}
H^{2}(a)/H_{0}^{2}&=&\Omega_{0b}a^{-3}+\Omega_{0r}a^{-4}+(1-\Omega_{0b}-\Omega_{0r})[A_{s}^{2}+(1-A_{s})^{2}a^{-3}+2A_{s}(1-A_{s})a^{-3/2}]\nonumber\\
&&=\Omega_{0b}a^{-3}+\Omega_{0r}a^{-4}+\Omega_{01}+\Omega_{02}a^{-3}+\Omega_{03}a^{-3/2},\label{H2-A0}
\end{eqnarray}
where $\Omega_{0b}$ and $\Omega_{0r}$ represent the fractional energy densities for baryon and radiation (including all relativistic particles, such as CMB photon $\Omega_{0\gamma}$, neutrino $\Omega_{0\nu}$, {\it etc.}.), respectively. From  Eq. (\ref{H2-A0}), one easily gets the current dimensionless energy density for the dark energy $\Omega_{\Lambda}=\Omega_{01}=(1-\Omega_{0b}-\Omega_{0r})A_{s}^{2}$, dark-matter $\Omega_{0dm}=\Omega_{02}=(1-\Omega_{0b}-\Omega_{0r})(1-A_{s})^{2}$ and unfound  component $\Omega_{0u}=\Omega_{03}=2(1-\Omega_{0b}-\Omega_{0r})A_{s}(1-A_{s})$.

  After calculation, one gains $A_{s}\in (0.39,0.6)$,  $\Omega_{\Lambda}\in (0.15,0.34)$  and $\Omega_{0u}\in(0.45,0.46)$ by setting  current values $\Omega_{0r}\sim 0$, $\Omega_{0b}=0.05$ and $\Omega_{0m}\in (0.2,0.4)$. It is obvious that  the value of DE density is smaller than observations due to the existence of $\Omega_{0u}$.  Taking $a=1$ in Eq. (\ref{H2-A0}), we have $\sqrt{\Omega_{\Lambda}}=\sqrt{1-\Omega_{0b}-\Omega_{0r}}-\sqrt{\Omega_{0dm}}$.  Via this relation, the values of $\Omega_{\Lambda}$ and $\Omega_{0m}$ are illustrated in Fig. \ref{cs2-omega-A=0},  where one can read in RMCG model the deviation of density-parameter values from $\Lambda$CDM. Furthermore, we can slove $A_{s}\simeq0.49$ and  $\Omega_{\Lambda}\simeq0.23$ when  we take $\Omega_{0m}=0.3$.
\begin{table}[!htbp]
 \vspace*{-12pt}
 \begin{center}
 \begin{tabular}{  | c| c| c |  c |} \hline\hline
        Density parameter        &  Explicit form    &   Parameter value     &  EOS  \\\hline
           $\Omega_{0r}$      &       $\Omega_{0r}$      &  ----   &   $w=1/3$ \\\hline
           $\Omega_{0b}$      &        $\Omega_{0b}$   &   $0.05$   &   $w=0$ \\\hline
             $\Omega_{0dm}$           &    $(1-\Omega_{0b}-\Omega_{0r})(1-A_{s})^{2}$     &     (0.15,0.35)       & $w=0$  \\\hline
          $\Omega_{\Lambda}$        &  $(1-\Omega_{0b}-\Omega_{0r})A_{s}^{2}$        &      (0.15,0.34)      &  $w=-1$  \\\hline
       $1-\Omega_{\Lambda}-\Omega_{0dm}-\Omega_{0b}-\Omega_{0r}$   & $(1-\Omega_{0b}-\Omega_{0r})2A_{s}(1-A_{s})$     &   (0.45,0.46)  & $w=-1/2$ \\\hline\hline
 \end{tabular}
 \end{center}
 \caption{Values of dimensionless  density parameters in RMCG ($A=0$) cosmology. }\label{table-omega-A=0}
 \end{table}

 Analyzing  the evolution of deceleration parameter $q(z)$, we find in Fig. \ref{cs2-omega-A=0} that cosmic expansion is translated from deceleration to acceleration, where the current value $q_{0}\in(-0.355,-0.056)$  given by the RMCG ($A=0$) model is  larger than  $q_{0}\in(-0.7,-0.4)$ given by the standard $\Lambda$CDM cosmology. For plotting Fig. \ref{cs2-omega-A=0} we use  the  parameter values $A_{s}=[0.39,0.49,0.6]$ corresponding to $\Omega_{0m}=[0.2,0.3,0.4]$,  respectively. From the evolution of  $w(z)$ plotted in Fig. \ref{cs2-omega-A=0}, one receives the result that the negative pressure is provided by the RMCG fluid at late time of our Universe. For  the evolutions of $c_{s}^{2}(z)$,  the unexpected negative sound speed is appeared in this RMCG fluid. Since this unified fluid  includes dust component, the negative sound speed will induce  the classical instability  to the system at structure form, where the perturbations on small scales will increase quickly with time and the late time history of the structure formations will be significantly modified \cite{cs-question}. Then it seems that the RMCG (A=0) model is not a good one.

\begin{figure}[ht]
   \includegraphics[width=4.3cm]{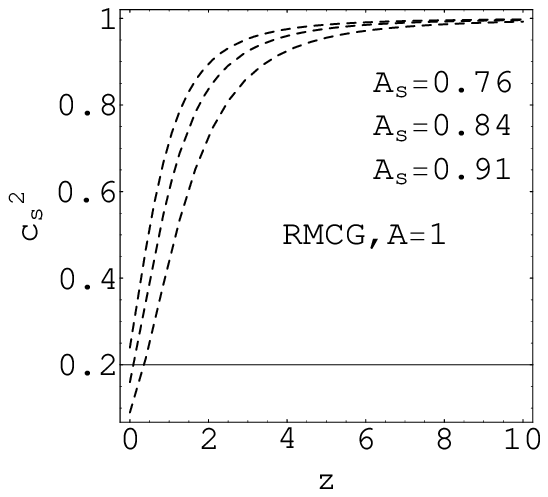}
    \includegraphics[width=4.3cm]{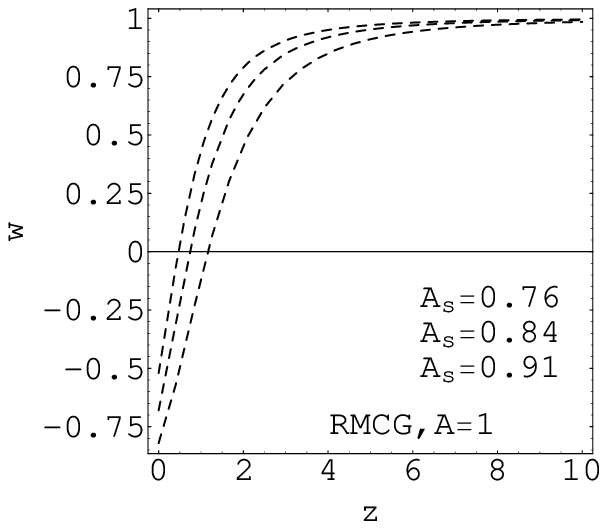}
  \includegraphics[width=4.3cm]{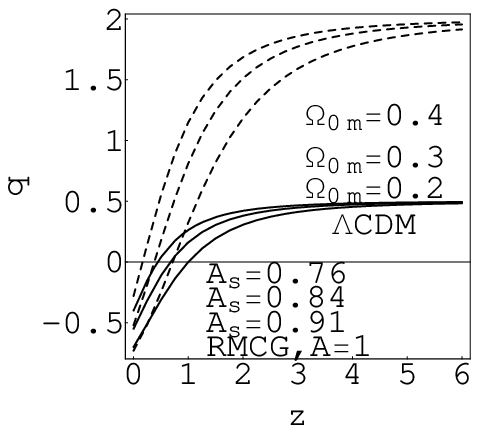}
   \includegraphics[width=4.3cm]{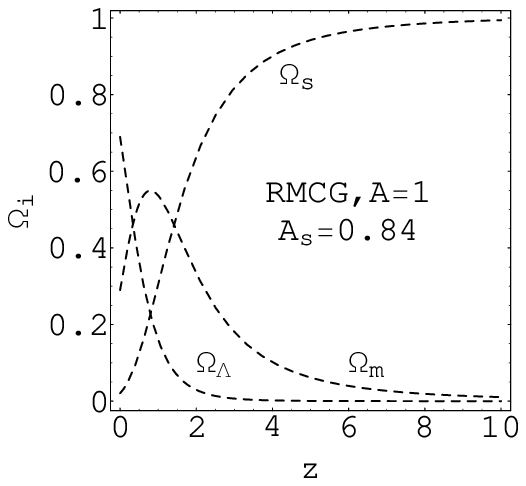}\\
  \caption{Evolutions of the $c_{s}^{2}$,  $w$,  $q$  and  $\Omega_{i}$ versus $z$ for the RMCG ($A=1$) model.}\label{cs2-q-w-A=1}
\end{figure}

For $A=1$,   RMCG fluid contains the DE, DM and stiff matter ($w=1$), where the Friedmann equation is expressed by
 \begin{eqnarray}
H^{2}(a)/H_{0}^{2}&=&\Omega_{0b}a^{-3}+\Omega_{0r}a^{-4}+(1-\Omega_{0b}-\Omega_{0r})[A_{s}^{2}+2A_{s}(1-A_{s})a^{-3}+(1-A_{s})^{2}a^{-6}]\nonumber\\
&&=\Omega_{0b}a^{-3}+\Omega_{0r}a^{-4}+\Omega_{01}+\Omega_{02}a^{-3}+\Omega_{03}a^{-6}.\label{H2-A1}
\end{eqnarray}
One from  Eq. (\ref{H2-A1}) gains $\Omega_{\Lambda}=\Omega_{01}=(1-\Omega_{0b}-\Omega_{0r})A_{s}^{2}$, $\Omega_{0dm}=\Omega_{02}=2(1-\Omega_{0b}-\Omega_{0r})A_{s}(1-A_{s})$ and $\Omega_{0s}=\Omega_{03}=(1-\Omega_{0b}-\Omega_{0r})(1-A_{s})^{2}$.
 Taking $\Omega_{0m}\in (0.2,0.4)$, we receive $A_{s}\in (0.76,0.91)$,  $\Omega_{\Lambda}\in (0.55,0.79)$ and $\Omega_{0s}\in(0.01,0.05)$, which are listed in table \ref{table-omega-A=1}. For this case,   $\Omega_{0m}=0.3$ gives $\Omega_{\Lambda}=0.67$.

\begin{table}[!htbp]
 \vspace*{-12pt}
 \begin{center}
 \begin{tabular}{  | c| c| c |  c |} \hline\hline
        Density parameter        &  Explicit form    &   Parameter value     &  EOS  \\\hline
           $\Omega_{0r}$      &       $\Omega_{0r}$      &  ----   &   $w=1/3$ \\\hline
           $\Omega_{0b}$      &        $\Omega_{0b}$   &   $0.05$   &   $w=0$ \\\hline
             $\Omega_{0dm}$           &    2$(1-\Omega_{0b}-\Omega_{0r})A_{s}(1-A_{s})$     &     (0.15,0.35)       & $w=0$  \\\hline
          $\Omega_{\Lambda}$        &  $(1-\Omega_{0b}-\Omega_{0r})A_{s}^{2}$        &      (0.55,0.79)      &  $w=-1$  \\\hline
       $1-\Omega_{\Lambda}-\Omega_{0dm}-\Omega_{0b}-\Omega_{0r}$   & $(1-\Omega_{0b}-\Omega_{0r})(1-A_{s})^{2}$     &   (0.01,0.05)  & $w=1$ \\\hline\hline
 \end{tabular}
 \end{center}
 \caption{Values of dimensionless  density parameters in RMCG ($A=1$) cosmology. }\label{table-omega-A=1}
 \end{table}

 Fig.\ref{cs2-q-w-A=1} illustrates the evolutions of the adiabatic sound speed, EoS,  deceleration parameter and dimensionless energy density in the RMCG $(A=1)$. As we can see,  the value of EoS is transited  from the positive to the negative. Correspondingly, a transition from decelerating-expansion universe to accelerating-expansion universe can be realized. Meanwhile, this  RMCG unified fluid at hand would not bring the negative value of the adiabatic sound speed. But it has other problems we have to face, such as (1)  deceleration parameter is $q>\frac{1}{2}$  at high redshift, which is not satisfied with $q\leq \frac{1}{2}$ in the  matter-dominate universe. Matter-dominate universe is necessary for structure formation; (2) Radiation-dominate universe will not appear in this RMCG universe, because of stiff matter. From these points, it seems that this model is not consistent with the current observational universe.

\subsection{$\text{ A unified model of dark energy and dark radiation  in RMCG cosmology}$}

Combined analysis of several cosmological data hints the existence of  an extra relativistic-energy component (called dark radiation) in the early universe,  in addition to the well-known three neutrino species predicted by the standard model of particle physics. The total amount of this extra DR component is  often related to the  parameter $N_{eff}$ denoting  the  effective  number  of relativistic degrees of freedom, which has relation to the energy density of relativistic  particles  via  $\rho_{\nu}=\frac{7}{8}(4/11)^{4/3}\rho_{\gamma}N_{eff}$. Here $\rho_{\nu}$ and $\rho_{\gamma}$ represent the fractional energy density for neutrino and  CMB photon, respectively. The entropy transfer between neutrinos and thermal bath modifies this number to $N_{eff} = 3.046$ \cite{DR-0612150,DR-11034132}. However, larger values of $N_{eff}$  are reported by the cosmic observations. Depending on the datasets, constraint results on $N_{eff}$  are qualitatively changed.   For instance,  it is pointed out that the observational deuterium abundance D/H favors the presence of extra radiation \cite{DR-DH1,DR-DH2}: $N_{eff} = 3.90 \pm 0.44$. The  combining analysis of CMB data from the 7-year WMAP and the Atacama Cosmology Telescope (ACT) gives an excess $N_{eff} = 5.3\pm1.3$ \cite{DR-12050553}, and the addition of BAO  and $H_{0}$ data decreases the value $N_{eff} = 4.56\pm0.75$  \cite{DR-12050553,DR-10090866}.  CMB data from the  9-year  WMAP combining with the South Pole Telescope (SPT) and  the 3-year Supernova Legacy Survey (SNLS3) provides  a non-standard value, $N_{eff} = 3.96 \pm 0.69$ \cite{DR-13030143,DR-13045243}. Ref. \cite{DR-13035076} shows that  $N_{eff}=3.62^{+0.50}_{-0.48}$ for using the Planck+WP+highL+$H_{0}$ and  $N_{eff} = 3.52^{+0.48}_{-0.45}$  for using the Planck+WP+highL+BAO+$H_{0}$, whose analysis suggests the presence of a dark radiation at 95\% confidence level. For more limits on $N_{eff}$, one can see Refs. \cite{DR-13070637,DR-11092767,DR-11053182}.

The above urgency to search  source of DR  is relieved by the study in Ref. \cite{DR-Hprior}. Given that $N_{eff}$ is degenerate with the value of $H_{0}$, Ref. \cite{DR-Hprior} focuses  on how the $H_{0}$ prior changes the value of $N_{eff}$, and obtains the result that a lower prior for $H_{0}$ moves the limits to lower $N_{eff}$. It is pointed out in Ref.   \cite{DR-Hprior} that there is no longer that much evidence supporting the existence of DR, since  this evidence is partially driven by the larger value $H_{0}=73.8 \pm 2.4 $ km s$^{-1}$Mpc$^{-1}$ from the HST while several measurements suggest the lower value of $H_{0}$, such as $H_{0}=68 \pm 2.8$ km s$^{-1}$Mpc$^{-1}$ from the median statistics (MS) analysis of the 537 non-CMB measurements \cite{H0MS}, $H_{0}=67.3 \pm 1.2$ km s$^{-1}$Mpc$^{-1}$ from the Planck+WP+highL \cite{DR-13035076} and $H_{0}=68.1 \pm 1.1$ km s$^{-1}$Mpc$^{-1}$ from the 6dF+SDSS+BOSS+WiggleZ BAO data sets  \cite{DR-13035076}. For model-dependent results, Ref. \cite{DR-Hprior} shows that in the $\Lambda$CDM it indicates the presence of DR  with the HST $H_{0}$ prior, while there is no significant statistical evidence for  existence of DR with the MS $H_{0}$ prior \cite{DR-Hprior}; In XCDM parametrization of time-evolving DE it brings the result: the evidence for DR is significant for both the HST $H_{0}$ prior and the MS $H_{0}$ prior  \cite{DR-Hprior}.

In this section, we explore the RMCG model that apparent extra DR directly links to the physics of the cosmological-constant (CC) DE. Fixing  $A=1/3$, RMCG fluid unifies the DE and DR, where the Friedmann equation becomes
\begin{eqnarray}
H^{2}(a)/H_{0}^{2}&=&\Omega_{0m}a^{-3}+(\Omega_{0\gamma}+\Omega_{0\nu})a^{-4}+
(1-\Omega_{0m}-\Omega_{0\gamma}-\Omega_{0\nu})[A_{s}^{2}+(1-A_{s})^{2}a^{-4}+2A_{s}(1-A_{s})a^{-2}]\nonumber\\
&&=\Omega_{0m}a^{-3}+(\Omega_{0\gamma}+\Omega_{0\nu})a^{-4}+\Omega_{01}+\Omega_{02}a^{-4}+\Omega_{03}a^{-2}.\label{H2-A3}
\end{eqnarray}
Here $\Omega_{01}=(1-\Omega_{0m}-\Omega_{0\gamma}-\Omega_{0\nu})A_{s}^{2}=\Omega_{\Lambda}$ is the energy density of cosmological-constant type DE,
$\Omega_{02}=(1-\Omega_{0m}-\Omega_{0\gamma}-\Omega_{0\nu})(1-A_{s})^{2}=\Omega_{0dr}$ is the coefficient of DR term that is a characteristic feature in the RMCG (A=$1/3$) fluid, the term $\Omega_{03}a^{-2}=2A_{s}(1-\Omega_{0m}-\Omega_{0\gamma}-\Omega_{0\nu})(1-A_{s})a^{-2}=\Omega_{0k}^{eff}a^{-2}$ dilutes as $a^{-2}$ jus like the curvature density in the non-flat geometry, called effective curvature density. In the non-flat universe, then the current curvature density is modified as $\Omega_{0k}+ \Omega_{0k}^{eff}$. Besides the RMCG fluid, we supplement the matter and radiation components in Eq. (\ref{H2-A3}).

  Eq. (\ref{H2-A3}) shows that the dimensionless density parameters (DE, DR and effective curvature density) relate to the  RMCG model parameter  $A_{s}$. The values of  these density parameters  should be consistent with observations. Given that relativistic particle includes the photon, neutrino and dark radiation, the total dimensionless density parameter of relativistic particle is written as $\Omega_{0r}^{tot}=\Omega_{0\gamma}+\Omega_{0\nu}+\Omega_{0dr}=\Omega_{0\gamma}[1+\frac{7}{8}(\frac{4}{11})^{4/3} N_{eff}]$, where the photon density parameter $\Omega_{0\gamma}=2.469\times 10^{-5}h^{-2}$ \cite{7ywmap}. Writing  $N_{eff}=N_{eff}^{SM}+\Delta N_{eff}$ and $ N_{eff}^{SM}=3.04$, one reads $\Omega_{0dr}=\frac{7}{8}(\frac{4}{11})^{4/3} \Omega_{0\gamma}\Delta N_{eff}$.  On the other hand, in the RMCG (A=$1/3$) model we receives $\Omega_{0dr}=\Omega_{02}=(1-\Omega_{0m}-\Omega_{0\gamma}-\Omega_{0\nu})(1-A_{s})^{2}$. Taking $\Omega_{0m}=0.3$  and    $\Delta N_{eff}=[0.5,1,2]$, we can calculate the values of  $A_{s}$ and the dimensionless density parameters, which are listed in table \ref{table-omega-A=033-Neff}. It is found from this table that the values of $\Omega_{\Lambda}$ and $\Omega_{0k}^{eff}$ are compatible to the cosmic observations \cite{9ywmap,DR-13035076},  where $\Omega_{\Lambda}$ is around 0.7 and  $\Omega_{0k}\sim 0$. And corresponding to $A_{s}<1$ (or $A_{s}>1$),  we have $\Omega_{0k}^{eff}>0$ (or $\Omega_{0k}^{eff}<0$).

  \begin{table}[!htbp]
 \vspace*{-12pt}
 \begin{center}
 \begin{tabular}{ |  c| c| c |  c | } \hline\hline
        $\Delta N_{eff}$    &  $A_{s}$    &   $\Omega_{\Lambda}$     &  $\Omega_{0k}^{eff}$  \\\hline
          0.5  &   0.9972 or  1.0028     &  0.6920 or 0.7080  &  0.0079 or -0.0080     \\\hline
          1    &   0.9960  or  1.0040     &  0.6944 or 0.7056  &  0.0056 or -0.0056         \\\hline
          2   &  0.9943 or 1.0057       &   0.6961 or 0.7039 &   0.0039 or -0.0039      \\\hline\hline
 \end{tabular}
 \end{center}
 \caption{ Values of $A_{s}$,  $\Omega_{\Lambda}$ and $\Omega_{0k}^{eff}$ calculated by using the values of $\Delta N_{eff}$ and fixing $\Omega_{0m}=0.3$. }\label{table-omega-A=033-Neff}
 \end{table}

\begin{figure}[ht]
      \includegraphics[width=4.3cm]{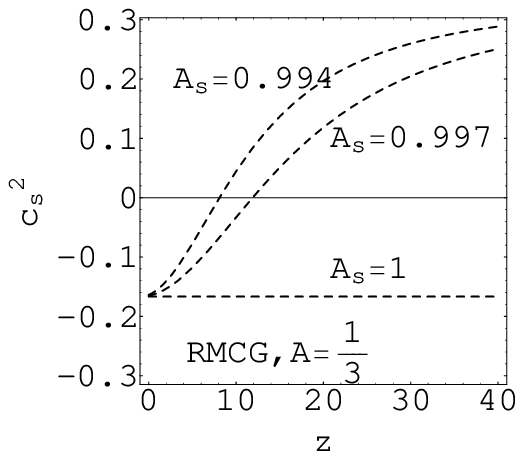}
    \includegraphics[width=4.3cm]{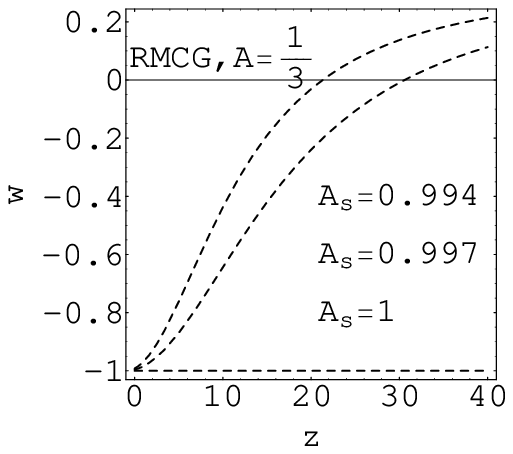}
       \includegraphics[width=4.3cm]{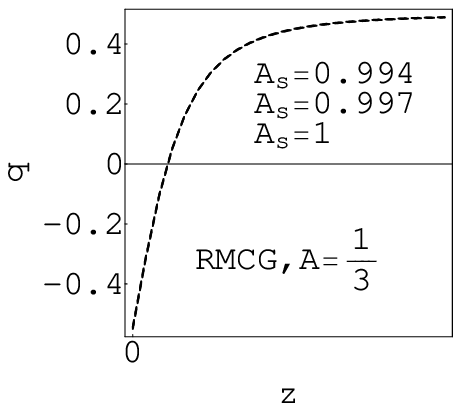}
     \includegraphics[width=4.3cm]{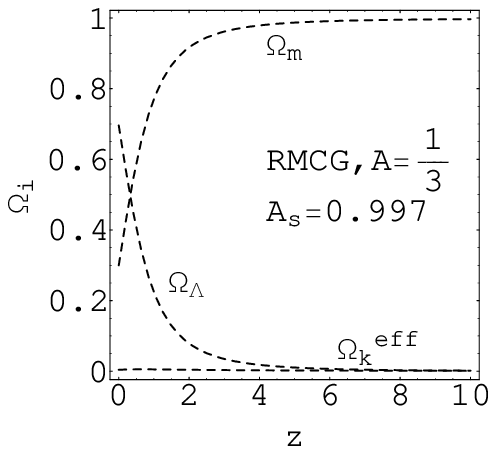}\\
  \caption{Behaviors of the $c_{s}^{2}$, $w$, $q$ and $\Omega_{i}$ versus $z$ for the RMCG ($A=1/3$) unified model of DE and DR.}\label{cs2-w-q-A=033}
\end{figure}

We plot pictures of the dimensionless density parameters and deceleration  parameter versus $z$. The third picture in Fig. \ref{cs2-w-q-A=033} describes a  universe having the transition from decelerated expansion  to accelerated expansion.  And the evolutions of $q(z)$ are almost the same for taking different value of $A_{s}$, due to a small variable region of  $A_{s}$ bounded by $\Delta  N_{eff}$. The values of current deceleration parameter and transition redshift are  $q_{0}=-0.546^{+0.004}_{-0.004}$ and $z_{T}=0.668^{+0.003}_{-0.004}$, a narrow range. Fig.\ref{cs2-w-q-A=033} also illustrates the evolution of $c_{s}^{2}(z)$ for the RMCG ($A=1/3$) fluid, where the positive value  of $c_{s}^{2}$ is converted  to the negative value with the evolution of universe. Since the RMCG ($A=1/3$) unified fluid do not include matter, the negative value of  $c_{s}^{2}$ will not destroy the structure  formation. Just as for the cosmological constant DE, we have $c_{s}^{2}=-1$.  The negative $c_{s}^{2}$ for DE is in fact necessary  if one requires the negative pressure to produce the accelerating universe.  This is not inconsistent with the structure formation. For the behavior of $w$, at late time we can get $w<0$ which can be responsibility to the accelerating universe, and  at early time we obtain $w\sim 1/3$. According to the analysis above, the behaviors of cosmological quantities in the RMCG ($A=1/3$) model are accordant with the current observational universe. Then the RMCG ($A=1/3$) model can be considered as a candidate for the DE and DR. At last, we note that we do not discuss the case of $A_{s}>1$ for the RMCG ($A=1/3$), since the $c_{s}^{2}$ and $w$  will be divergent  at some points (when $A_{s}=-(1-A_{s})a^{-\frac{3}{2}(1+A)}$).

\subsection{$\text{ RMCG fluid as  dark energy}$}

The unification of the DE and DM (or DR) have been discussed in above parts. In the following, we investigate other possible properties of the RMCG fluid by taking values of  $A$ (except $A= 0, 1$ and $1/3$). For $A>0$ ($A\neq 0, 1, 1/3$),  Eq. (\ref{density-EU}) states that the RMCG fluid contains the CC and other positive-pressure  or negative-pressure  components (depending on the concrete value of $A$). We know nothing about these indeterminate components, such as  their function in universe or their responsibility to observations. So, here we do not discuss the case of $A>0$.  For  $A<0$, the RMCG fluid plays a role as the dynamical phantom or dynamical quintessence DE, where the Friedmann equation is written as
\begin{eqnarray}
H^{2}(a)/H_{0}^{2}&=&\Omega_{0m}a^{-3}+\Omega_{0r}a^{-4}+\Omega_{0RMCG}[A_{s}^{2}+(1-A_{s})^{2}a^{-3(1+A)}+2A_{s}(1-A_{s})a^{\frac{-3(1+A)}{2}}]\nonumber\\
&&=\Omega_{0m}a^{-3}+\Omega_{0r}a^{-4}+\Omega_{01}+\Omega_{02}a^{-3(1+w_{2})}+\Omega_{03}a^{-3(1+w_{3})},\label{H2-A4}
\end{eqnarray}
where $\Omega_{0RMCG}=1-\Omega_{0m}-\Omega_{0r}$, $\Omega_{01}=\Omega_{0RMCG}A_{s}^{2}$, $\Omega_{02}=\Omega_{0RMCG}(1-A_{s})^{2}$ and $\Omega_{03}=2\Omega_{0RMCG}A_{s}(1-A_{s})$.  For $A<-1$, we easily get $w_{2}=A<-1$ and $w_{3}=\frac{A-1}{2}<-1$. So, the RMCG fluid comprises the CC and  phantom DE, which plays  a role as the phantom-type DE; For $0>A>-1$, the RMCG fluid includes the CC and quintessence DE; For $A=-1$ or  $A_{s}=1$, the RMCG fluid reduces to the CC. Since we in theory have $0\leq A_{s}\leq 1$  due to the constraint on the current dimensionless density parameter $0<\Omega_{0j}<1$,  we can get the limit $-1<A<-1/3$ for the quintessence-type DE; We can obtain the theoretical limit $A<-1$ with $ 0<A_{s}<1$  for the phantom-type DE. For $A_{s}>1$,  the phantom-type DE ($-1<A<0$) and the quintessence-type DE  ($A<-1$) are non-physical, which should be ruled out.

\begin{figure}[ht]
   \includegraphics[width=4.3cm]{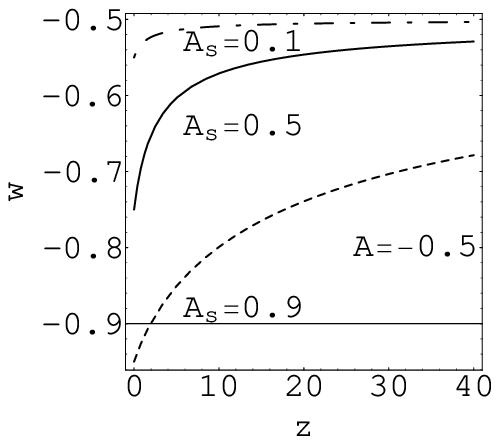}
   \includegraphics[width=4.3cm]{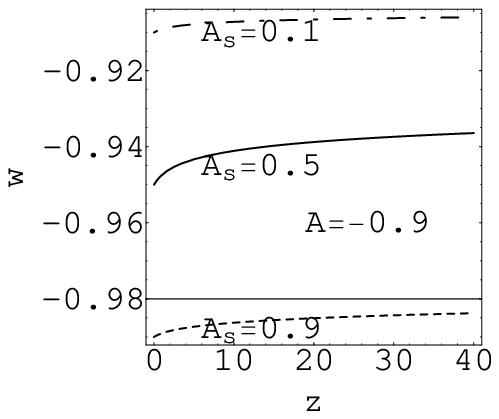}
   \includegraphics[width=4.3cm]{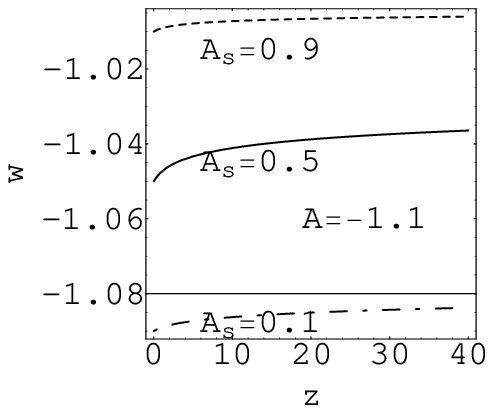}
   \includegraphics[width=4.3cm]{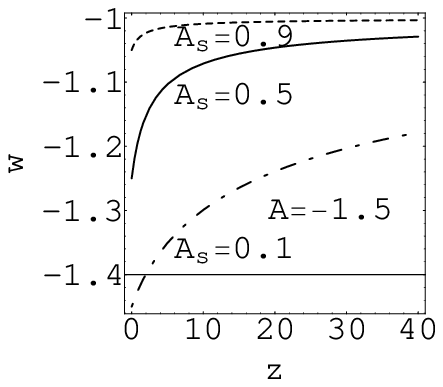}\\
    \includegraphics[width=4.3cm]{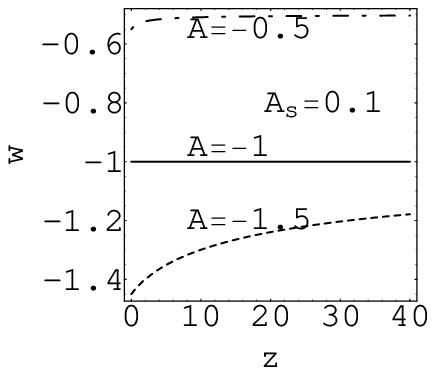}
    \includegraphics[width=4.3cm]{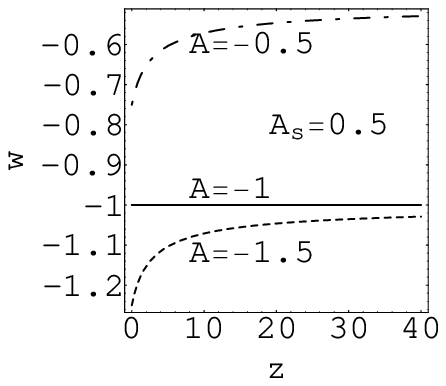}
   \includegraphics[width=4.3cm]{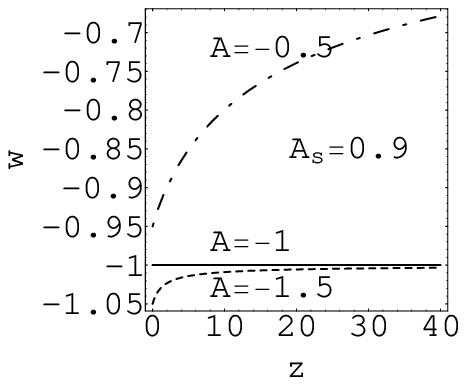}
   \includegraphics[width=4.3cm]{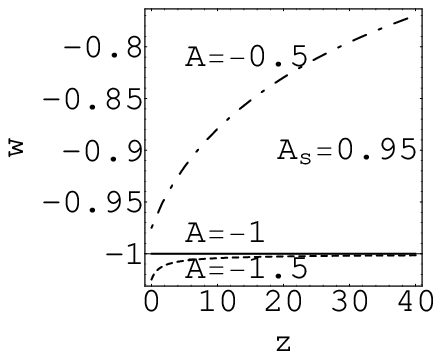}\\
  \caption{ Evolutions of $w(z)$ for the RMCG ($A<0$) fluid by taking  different  values of model parameters.}\label{w-A-negative}
\end{figure}

  Fig. \ref{w-A-negative} illustrates the dependence  of $w(z)$ on model parameters for the RMCG ($A<0$) fluid. From  Fig. \ref{w-A-negative}, we can read properties of $w(z)$. (1) The CC,  quintessence and  phantom DE can be realized in this RMCG fluid by taking different values of $A$ and $A_{s}$; (2) According to  four upper figures in Fig. \ref{w-A-negative},  for phantom (two upper-right  figures)  we have the result  that the less values  of parameters $A$ and $A_{s}$, the less value of $w$. For quintessence (two upper-left figures)  we have the results that the less value  of parameter  $A$,  the less value of $w$, while the less value  of parameter  $A_{s}$, the larger value of $w$; (3) As we can see from  four upper figures in Fig.\ref{w-A-negative},  the value of more near to $A=-1$, the less influence on $w$  from $A_{s}$.  Also, from  four lower figures in Fig. \ref{w-A-negative}, we obtain the result that the value of more near to $A_{s}=1$, the less influence on $w$  from $A$.

\begin{figure}[ht]
  \includegraphics[width=4.3cm]{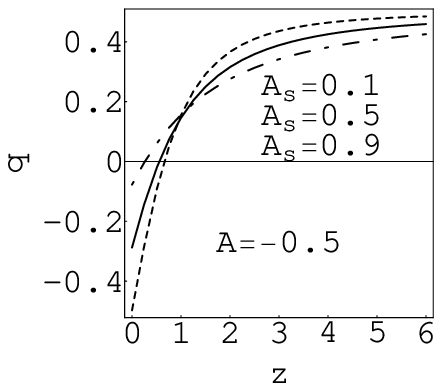}
   \includegraphics[width=4.3cm]{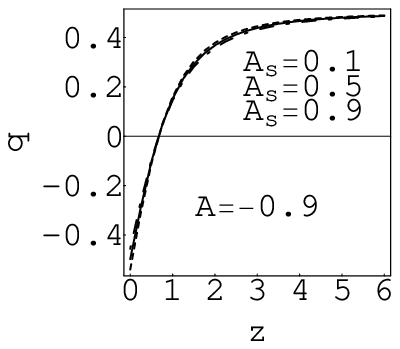}
   \includegraphics[width=4.3cm]{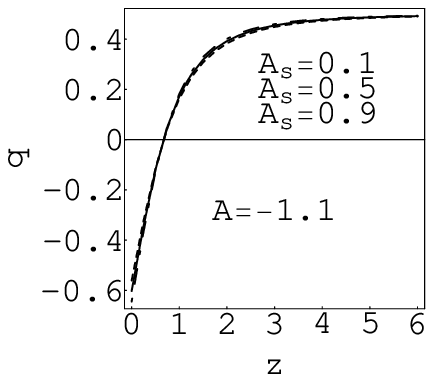}
   \includegraphics[width=4.3cm]{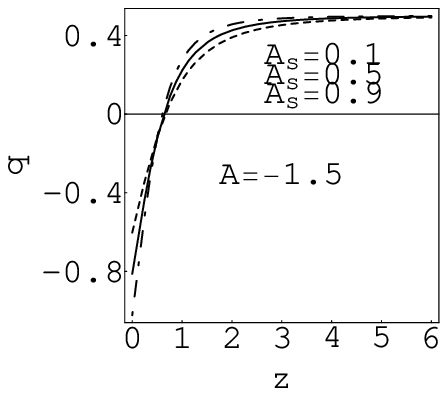}\\
    \includegraphics[width=4.3cm]{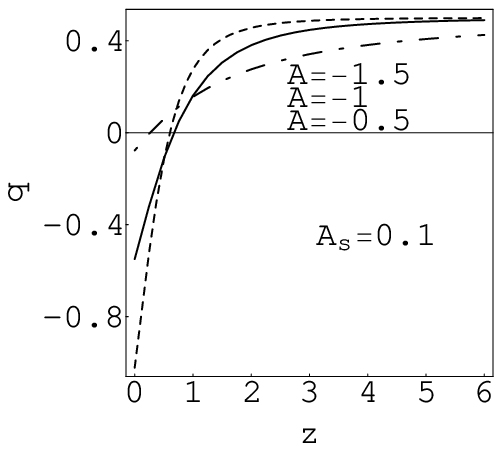}
     \includegraphics[width=4.3cm]{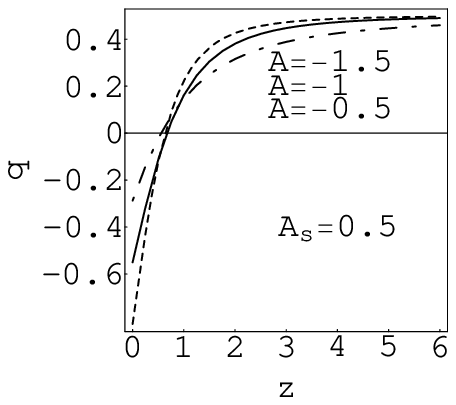}
   \includegraphics[width=4.3cm]{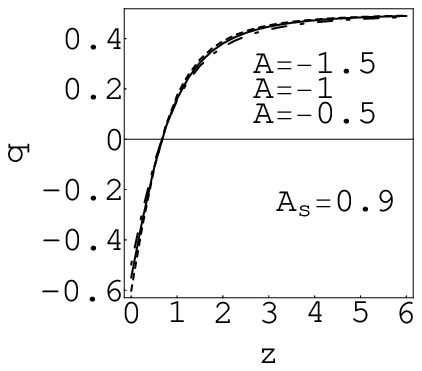}
   \includegraphics[width=4.3cm]{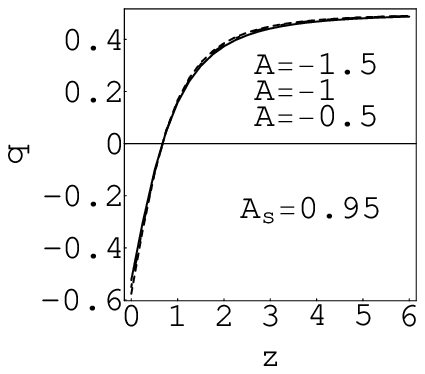}\\
  \caption{ Evolutions of $q(z)$ for the RMCG ($A<0$) model by taking different values of $A$ and $A_{s}$.}\label{q}
\end{figure}

Trajectories of $q(z)$ in the RMCG ($A<0$) model are drew in Fig. \ref{q}, which describe a universe transiting  from decelerating expansion to accelerating expansion. One  can also see  an interesting property for $q(z)$ from Fig. \ref{q}. The behavior of $q(z)$ is almost the same for using the different value of $A_{s}$ (or  $A$),  when the value of another model parameter $A$ (or $A_{s}$) is near to $-1$ (or $1$). For example,  $q(z)$ is not  sensitive to the change of value for $A_{s}$ (or  $A$), when we take $A=-0.9$ and $A=-1.1$ (or, $A_{s}=0.9$ and $A_{s}=0.95$). By the way, Fig. \ref{cs2-A-negative} illustrates the evolutions of $c_{s}^{2}(z)$ for RMCG ($A<0$) fluid, where the negative $c_{s}^{2}$ is obtained.

\begin{figure}[ht]
  \includegraphics[width=4.3cm]{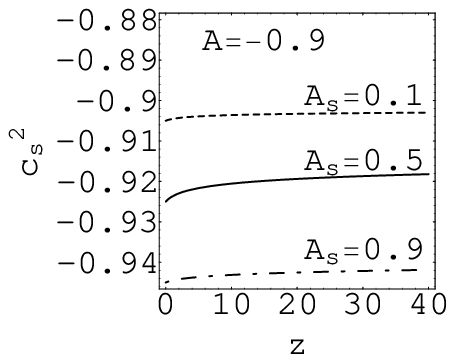}
  \includegraphics[width=4.3cm]{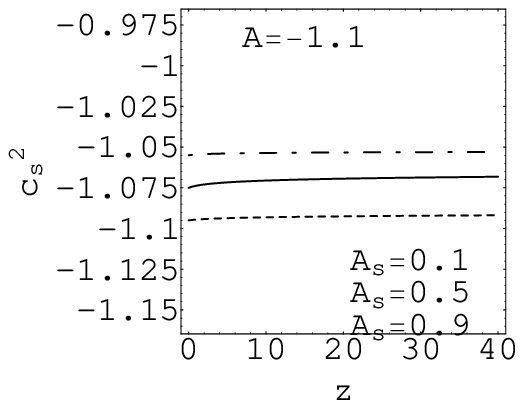}
   \includegraphics[width=4.3cm]{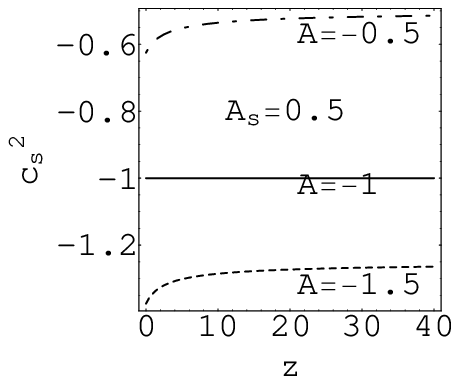}
  \includegraphics[width=4.3cm]{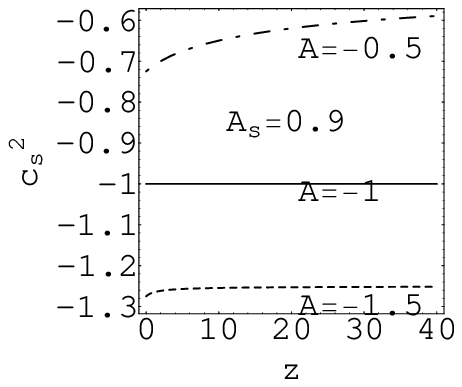}\\
  \caption{ Evolutions of the $c_{s}^{2}(z)$ for the RMCG ($A<0$) fluid by taking  different  values of model parameters.}\label{cs2-A-negative}
\end{figure}

\section{$\text{ Evolutions of  growth factor and Hubble parameter in the RMCG and comparisons with  cosmic data}$}

\begin{table}[ht]
\begin{center}
\begin{tabular}{|c||l|l|l|l|l|l|l|l|l|l|l|l|l|l|l|l|l|l|l|l|}\hline\hline
Number           & 1            & 2           & 3          & 4       & 5        & 6         &7         & 8           & 9        & 10    \\ \hline
$z$              & 0.15         & 0.22        & 0.32       & 0.35    & 0.41     & 0.55      &0.60      & 0.77        & 0.78     &1.4   \\ \hline
$f$              &0.51          &0.6          & 0.654      & 0.7     &0.7        & 0.75     &0.73      & 0.91         & 0.7     &0.9   \\ \hline
$\sigma$         &0.11  & 0.1    &0.18    & 0.18  & 0.07   &0.18  & 0.07  & 0.36   & 0.08 &0.24  \\ \hline
Ref.              & \cite{fdata,fdata1}     & \cite{fdata2}     & \cite{fdata3}       & \cite{fdata4}    & \cite{fdata2}
                 & \cite{fdata5}            &\cite{fdata2}      & \cite{fdata6}       &\cite{fdata2}     &\cite{fdata7}   \\ \hline\hline
\end{tabular}
\end{center}
\caption{\label{table-fdata} Data of  growth factor $f$  with errors at different redshift.}
\end{table}

\begin{table}[ht]
\begin{center}
\begin{tabular}{|c||l|l|l|l|l|l|l|l|l|l|l|l|l|l|l|l|l|l|l|l|l|l|l|l|l|l|}\hline\hline
Number    & 1    & 2      & 3     & 4     & 5      & 6       &7       & 8     & 9      & 10    & 11   & 12    &13   &14   &15    \\ \hline
$z$       &0.07  & 0.09   &0.10   & 0.12  & 0.17   & 0.179   &0.199   & 0.2   & 0.27   &0.28   &0.35  & 0.352 &0.40 &0.44 &0.48  \\ \hline
$H$       &69    & 69     &69     & 68.6  & 83     & 75      &75      &72.9   & 77     &88.8   &76.3  & 83    &95   &82.6 &97    \\ \hline
$\sigma$  &19.6  & 12     &12     & 26.2  & 8      & 4       &5       &29.6   & 14     &36.6   &5.6   & 14    &17   &7.8  &62    \\ \hline
Ref.      &\cite{Hdata5}  & \cite{Hdata1}    & \cite{Hdata1}  & \cite{Hdata5}  & \cite{Hdata1}  &\cite{Hdata3} & \cite{Hdata3} &\cite{Hdata5}
          & \cite{Hdata1} & \cite{Hdata5}    &\cite{Hdata7}  & \cite{Hdata3}   & \cite{Hdata1}    &\cite{Hdata6}  & \cite{Hdata2}\\ \hline\hline
Number   &16    & 17  & 18   & 19   & 20    & 21    & 22   &23    & 24    & 25   & 26   & 27   & 28   &29 &   \\ \hline
$z$      &0.593 &0.6  &0.68  &0.73  &0.781  &0.875  &0.88  &0.90  &1.037  &1.30  &1.43  &1.53  &1.75  &2.3  & \\ \hline
$H$      &104   &87.9 &92    &97.3  &105    &125    &90    &117   &154    &168   &177   &140   &202   &224  & \\ \hline
$\sigma$ &13    &6.1  &8     &7.0   &12     &17     &40    &23    &20     &17    &18    &14    &40    &8    & \\ \hline
Ref.     &\cite{Hdata3}  & \cite{Hdata6}    & \cite{Hdata3}  & \cite{Hdata6}  & \cite{Hdata3}  &\cite{Hdata3} & \cite{Hdata2} &\cite{Hdata1}
         & \cite{Hdata3} & \cite{Hdata1} &\cite{Hdata1} & \cite{Hdata1}   &\cite{Hdata1} & \cite{Hdata4}&\\ \hline\hline
\end{tabular}
\end{center}
\caption{\label{table-29Hubbledata} $H(z)$ data  with errors at different redshift (in units [${\rm km~s^{-1}\,Mpc^{-1}}$]).}
\end{table}

\begin{figure}[ht]
   \includegraphics[width=7.5cm]{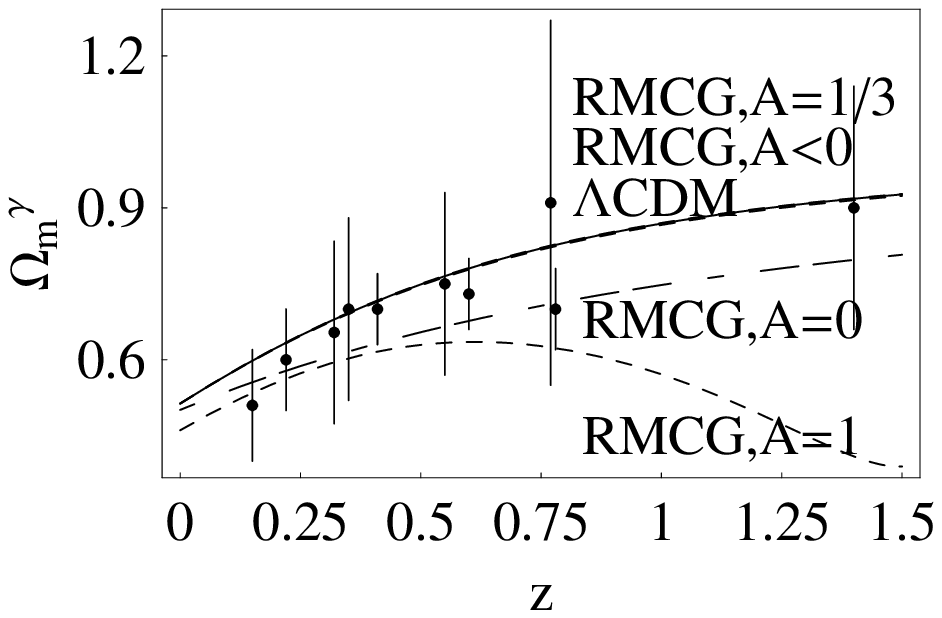}
   \includegraphics[width=7.5cm]{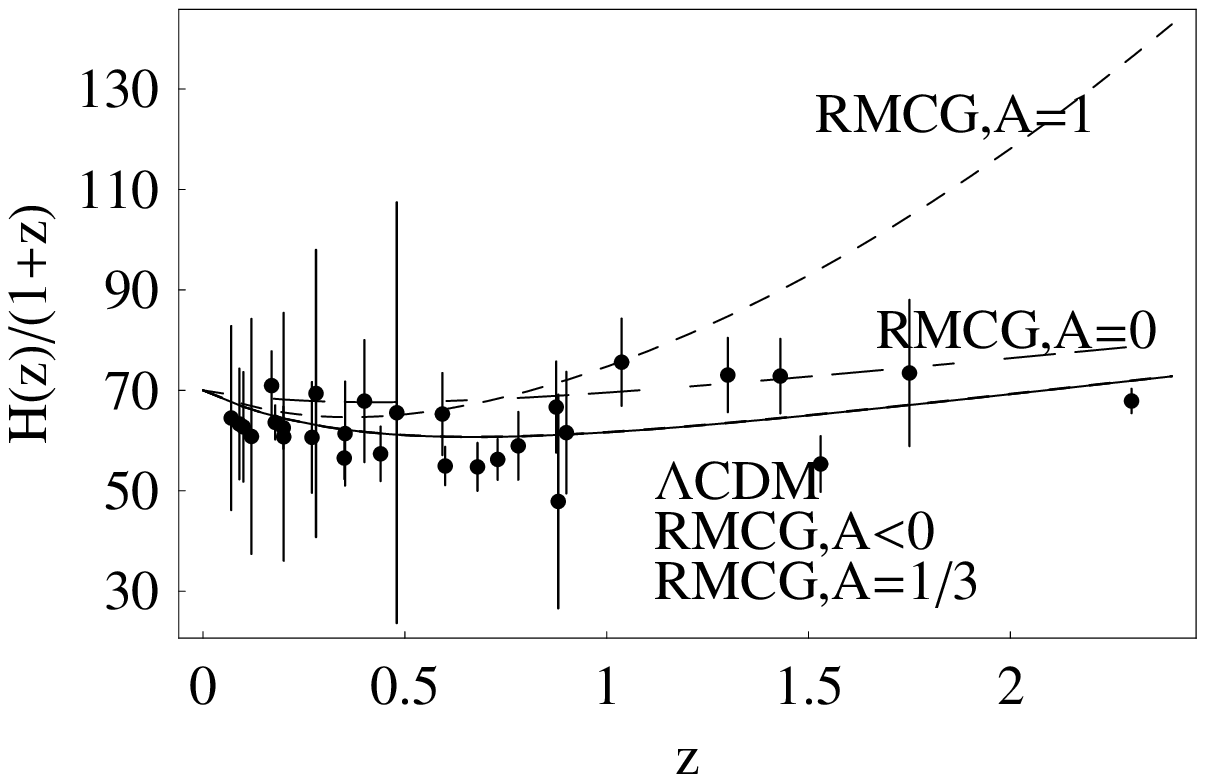}\\
  \caption{Evolutions of $\Omega_{m}^{\gamma}(z)$ and $H(z)/(1+z)$   versus $z$ for the RMCG and $\Lambda$CDM model.}\label{f-A=1}
\end{figure}

Via the cosmic observations, peoples  obtain some values of growth factor $f$ \cite{fdata,fdata1,fdata2,fdata3,fdata4,fdata5,fdata6,fdata7} and Hubble parameter $H$ \cite{Hdata1,Hdata2,Hdata3,Hdata4,Hdata5,Hdata6,Hdata7}, which are listed in table \ref{table-fdata} and \ref{table-29Hubbledata}.
 We apply the $f$ and  $H$ to test the RMCG models by comparing them with the observational data. Growth factor is defined as $f\equiv d\ln\delta/d\ln a$, which complies  with the following  equation
\begin{equation}
\frac{df}{da} +\frac{f^{2}}{a}+[\frac{2}{a}+\frac{(d\ln
H)}{da}]f-\frac{3\Omega_{m}(a)}{2}=0,\label{f2}
\end{equation}
deriving by the perturbation equation $\ddot{\delta}+2H\dot{\delta}-4\pi G \rho_{m}\delta=0$. Here $\delta\equiv \delta \rho_{m}/\rho_{m}$ is the matter density contrast  and  "dot" denotes the derivative with respect to cosmic time $t$. Usually , it is hard to find the analytical solutions to Eq. (\ref{f2}). The approximation $f\simeq \Omega_{m}^{\gamma}$ has been used in many papers,  which  provides an excellent  fit to the numerical form of $f(z)$   for various cosmological models  \cite{f-form,f-form1,f-form2,f-form3,f-form4,f-form5}. Growth index $\gamma$ can be given by considering the zeroth order and the first order terms in the expansion for $\gamma$ \cite{f-gamma}, $\gamma=\frac{3(1-w)}{(5-6w)}+\frac{3(1-w)(1-\frac{3}{2}w)(1-\Omega_{m})}{125(1-\frac{6w}{5})^{3}}$. We illustrate the $\Omega_{m}^{\gamma}$ versus $z$ in Fig. \ref{f-A=1} by taking $\Omega_{0m}=0.3$ and $A_{s}=0.49$ for the RMCG ($A=0$), $\Omega_{0m}=0.3$ and $A_{s}=0.84$ for the RMCG ($A=1$),  $\Omega_{0m}=0.3$ and $A_{s}=0.997$ for the RMCG ($A=1/3$), $\Omega_{0m}=0.3$, $A_{s}=0.95$ and $A=-1.1$ for the RMCG ($A<0$).  It can be seen from Fig.\ref{f-A=1} that the  behaviors of $\Omega_{m}^{\gamma}(z)$ in the RMCG ($A=1/3$) and  RMCG ($A<0$) model  are almost the same as  the popular $\Lambda$CDM model (solid line in Fig.\ref{f-A=1}), where an  increasing function versus $z$   is consistent with the current observations. But, $\Omega_{m}^{\gamma}(z)$ in the RMCG ($A=1$) much deviates from that in the $\Lambda$CDM model at the higher redshift. For clarity,  we plot the trajectories of $H(z)/(1+z)$ for the discussional models,  and compare them  with the 29 observational $H(z)$ data listed in table \ref{table-29Hubbledata}. The difference of pictures between the RMCG ($A=1$) model and $\Lambda$CDM model is apparent at high redshift. And at the high redshift, the evolutions of $\Omega_{m}^{\gamma}(z)$ and $H(z)/(1+z)$  in the RMCG  ($A=1$) obviously deviate from the  observational data. From above, it is shown that the RMCG  ($A=1$) fluid as the unification of dark matter and dark energy is not well accordant with the  $f$ data and the Hubble data. But,  the RMCG ($A=1/3$) and RMCG ($A<0$) model are well consistent with these two cosmic datasets.

\section{$\text{Parameter evaluation and model comparison}$}

\begin{figure}[ht]
  \includegraphics[width=6cm]{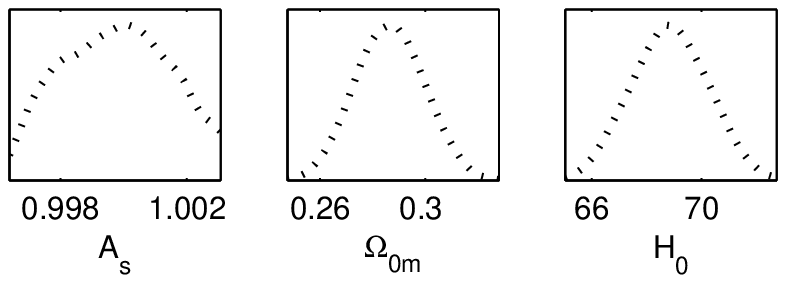}
  \includegraphics[width=10cm]{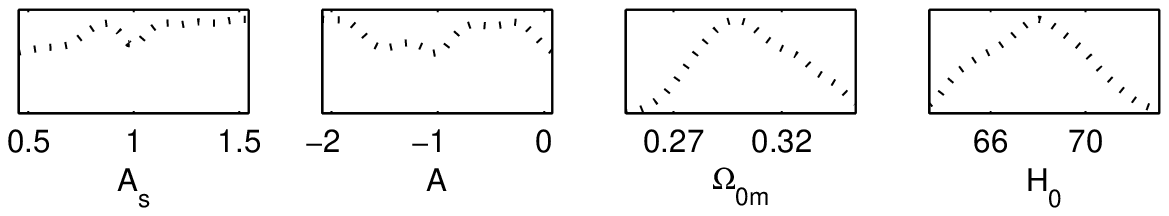}\\
  \caption{The 1D distribution of model parameters for the RMCG1  (left) and RMCG2 (right)  model.}\label{rmcg}
\end{figure}

\begin{figure}[ht]
   \includegraphics[width=5cm]{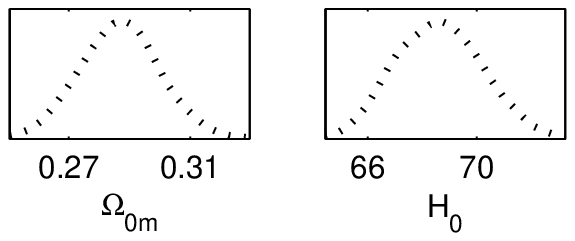}
  \includegraphics[width=11.5cm]{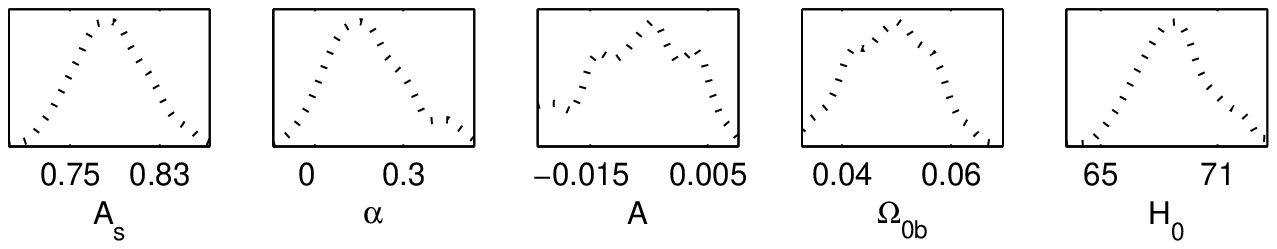}\\
  \caption{The 1D distribution of model parameters for the $\Lambda$CDM (left) and MCG model (right).}\label{mcg}
\end{figure}

In this section, we investigate the parameter space of the RMCG model. It can be known from the analysis above that  the RMCG unified model of the DE and DM are not favored, which have some questions on structure formation. For the RMCG (A=0) unified model, a negative sound speed will introduce the  instability at structure formation.  For the RMCG (A=1) unified model, perturbation quantity $f$ is not compatible with cosmic data, and a super-deceleration ($q>\frac{1}{2}$) expanded universe is not satisfied with the matter-dominate universe. So, these two cases will not studied in the following. We discuss the cosmic constraint on  the RMCG models with $A=1/3$ (RMCG1) and $A<0$ (RMCG2).  The data we use includes: baryon acoustic oscillation (BAO) data from the WiggleZ \cite{BAO-WiggleZ}, 2dfGRs \cite{BAO-2dFGRs} and SDSS \cite{BAO-SDSS} survey,  X-ray cluster gas mass fraction \cite{MCMC-fgas-data},  Union2 dataset of type supernovae Ia (SNIa) \cite{557Union2}  and 29 Hubble data listed in table \ref{table-29Hubbledata}. The constraint methods are described in Appendix.    For RMCG1, we have  $A_{s}=0.9993^{+0.0016+0.0028}_{-0.0016-0.0028}$, $\Omega_{0m}=0.287^{+0.012+0.024}_{-0.012-0.024}$  and $H_{0}=68.84^{+1.32+2.65}_{-1.32-2.47}$ with 68\% and 95\%  confidence levels. Obviously, $A_{s}$ is near to 1 and has the small confidence level. This calculation result for $A_{s}$  is approximatively equal to the cosmic constraint on $\Delta N_{eff}\in (0,1)$, which is consistent with other combining  constraints on $N_{eff}$ \cite{DR-13035076}. By the analysis of error-propagation, we calculate the DE density $\Omega_{\Lambda}=0.713^{+0.012+0.024}_{-0.012-0.024}$. For RMCG2, we find that $\Omega_{0m}=0.297^{+0.015+0.031}_{-0.016-0.028}$ and $H_{0}=68.25^{+1.46+2.92}_{-1.45-3.02}$, while the model parameters $A$  and $A_{s}$ are not convergent. The results are illustrated in Fig. \ref{rmcg}. From Eq. (\ref{H2-A4}),  we notice tha RMCG2 DE model reduces to the popular CC model by fixing $A=-1$ (or $A_{s}=1$), whatever value of $A_{s}$ (or $A$) is taken. The  non-convergent results on $A$  and $A_{s}$ may be  interpreted that the RMCG2 model can not be distinguished from the CC model by the cosmic data used in this paper.

\begin{table}
 \vspace*{-12pt}
 \begin{center}
 \begin{tabular}{|c |c | c |c |c |} \hline
 Case model               &  Free parameters                                          & $\chi^{2}_{min}$~~~     & $K$   & $\Delta AIC$  \\ \hline
 RMCG1 ($A=\frac{1}{3}$)  &  $\Omega_{0m}$, $A_{s}$,  $H_{0}$, $\Omega_{0r}$                   & 604.976 ~~~             &  3   & 0     \\ \hline
 $\Lambda$CDM             &$\Omega_{0m}$, $H_{0}$, $\Omega_{0r}$, $\Omega_{0dr}$               & 604.979        ~~~      &  3    & 0.003   \\  \hline
 RMCG2 ($A<0$)            &  $\Omega_{0m}$, $A_{s}$, $A$,  $H_{0}$, $\Omega_{0r}$, $\Omega_{0dr}$  & 603.636     ~~~        &  5   & 2.660   \\ \hline
 MCG            &  $\Omega_{0b}$, $A_{s}$, $A$, $\alpha$, $H_{0}$, $\Omega_{0r}$, $\Omega_{0dr}$   & 603.243     ~~~        &  6   & 4.267   \\ \hline
 \end{tabular}
 \end{center}
 \caption{Information criteria results. } \label{AICBIC}
 \end{table}

Next we use the objective information criteria (IC) to estimate the quality of above RMCG models. Akaike information criteria (AIC) is defined as \cite{AICselection,AICselection1}
\begin{equation}
AIC=-2\ln {\cal L}_{max}+2K,\label{AIC}
\end{equation}
where ${\cal L}_{max}$ is the highest likelihood in the model  with  $-2\ln {\cal L}_{max}=\chi^{2}_{min}$, $K$ is the number of free parameters to interpret the complexity of model. Usually, candidate model which minimizes the AIC is usually  considered the best. Comparing with the best one, one can calculate the difference for other model $\Delta$AIC $ = \Delta \chi_{min}^{2} + 2\Delta K$.  The rules for judging the strength of models are as follows. For $0\leq$ $\Delta$AIC$_{i}$$\leq 2$, model $i$  almost gains the same data support as the best model; for $2\leq $ $\Delta$AIC$_{i}$$\leq 4$, model $i$ gets the less support; and with $\Delta$AIC$_{i}$$ >10$ model $i$ is practically irrelevant  \cite{AICselection}.

 Since several observations imply the existence of DR, we take the DR density $\Omega_{0dr}$ as an additional free parameter in the $\Lambda$CDM, RMCG2 and MCG models. But, $\Omega_{0dr}$ is naturally included  in the RMCG1 model by the relation between $\Omega_{0dr}$ and model parameter $A_{s}$ and $\Omega_{0m}$. According to the calculation results in  table \ref{AICBIC}, one reads that the best model is the RMCG1. But, the $\Lambda$CDM model almost receives the same support as the RMCG1, since they  almost have the same AIC values. Comparing with the best RMCG1 model, the $\Delta$AIC values of the RMCG2 and MCG model  are calculated, too. From table \ref{AICBIC}, it is easy to see that the  RMCG2 model  is less supported by the AIC model-selection method, since $\Delta AIC=2.660$ at the range from 2 to 4. In addition, though the MCG model has the minimum value of $\chi^{2}$, it is  not favored  by analysis of the AIC, as it has the more large value $\Delta AIC=4.267$ resulted by the more model parameters. Corresponding to the  $\chi^{2}_{min}$ value, the constraint results on free parameters are $\Omega_{0m}=0.286^{+0.012+0.024}_{-0.012-0.023}$ and  $H_{0}=68.57^{+1.31+2.60}_{-1.31-2.43}$  for the $\Lambda$CDM model;  $A_{s}=0.788^{+0.031+0.060}_{-0.028-0.063}$, $\alpha=0.167^{+0.121+0.236}_{-0.110-0.205}$, $A=-0.0041^{+0.0063+0.0102}_{-0.0060-0.0139}$, $\Omega_{0b}=0.0501^{+0.0090+0.0160}_{-0.0093-0.0173}$ and $H_{0}=68.46^{+1.55+2.87}_{-1.44-3.01}$ for the MCG model. Using the best-fit model parameters and the covariance matrix, we find that all the four models listed in table \ref{zda}  show the  presence of a cosmological deceleration-acceleration transition.  The best-fit values  of translation redshift $z_{da}$  are 0.70, 0.70, 0.67 and 0.69  corresponding to the RMCG1, $\Lambda$CDM, RMCG2 and MCG model, respectively.  The mean with standard deviation are $0.71\pm 0.03$, $0.71\pm 0.03$,  $0.68\pm 0.03$ and $0.68\pm 0.05$ corresponding to the RMCG1, $\Lambda$CDM, RMCG2 and MCG model, respectively.  These values are in agreement with the result $z_{da}=0.74\pm 0.05$ given by Ref. \cite{z-da}.

\begin{table}
 \vspace*{-12pt}
 \begin{center}
 \begin{tabular}{|c |c | c |} \hline
 Case model                   &Best-fit $z_{da}$ & Mean  with standard deviation  \\ \hline
 RMCG1 ($A=\frac{1}{3}$)      & 0.70     &  $0.71\pm 0.03$      \\ \hline
 $\Lambda$CDM                 & 0.70     &  $0.71\pm 0.03$      \\  \hline
 RMCG2 ($A<0$)                & 0.67     &  $0.68\pm 0.03$      \\ \hline
 MCG                          & 0.69     &  $0.68\pm 0.05$      \\ \hline
 \end{tabular}
 \end{center}
 \caption{Values of cosmological deceleration-acceleration transition redshift $z_{da}$. } \label{zda}
 \end{table}

 One can notice that  the other criticism mechanism----Bayesian information criteria (BIC) that is defined as $BIC=-2\ln {\cal L}_{max}+K\ln n$ \cite{BICselection} is not studied in this paper. Here $n$ is the number of datapoints in the fitting. As we can see from the BIC definition,  the BIC value not only depends on the number of free parameter $K$ and the value of $\chi^{2}$, but also depends on the number  of datapoints $n$. So,   for the same models the different evaluation results would be given by the BIC analysis (induced by the different values of $\ln n$)   when one uses the different datapoints. For instance,  the value of $\ln n$ is obviously different for case of including  or not including SNIa data in combining constraint, since the SNIa data have the large number.   Given that the datapoint are always increasing, it seems that the calculation result from BIC is not "fair" for  more-parameter model when the more datapoints are given.  Quantitatively, the AIC and BIC method can give the   same result for $\ln n =2$ ($n\simeq 7.4$).   For datapoints  used in our analysis, it has $\ln n =6.452$. Seeing that the BIC is not "absolutely objective", i.e. its value much depends on the number of datapoints one use,  here we do not apply the BIC criticism method to evaluate the RMCG models.

\section{$\text{Conclusions}$}

  The RMCG models are  from the subclass of  the famous  MCG model that has been studied in great detail over the years. But, most of them were studied as a unification of DM and DE in the past. In this paper, we study the RMCG cosmology from a different point of view. We discuss the different cases in which the RMCG is regarded as the DE or the unified model. New interesting physical results are obtained in the RMCG dark models. The results show that (1) the RMCG  unified model of the dark energy and dark matter (with model parameter $A=0$ or $A=1$) tends to be ruled out by analysing  the behaviors of cosmological quantities.  For example, the RMCG (A=0) unified model appears a negative sound speed  which leads to the instability of the structure formation,  growth factor $f$  in the RMCG (A=1) unified model is not consistent with cosmic observational data. In addition,  a super-deceleration expanded universe ($q>1/2$) is not satisfied at the matter-dominate epoch and a radiation-dominate universe will not appear  in the RMCG  (A=1) model, due to the stiff matter; (2) the RMCG ($A=1/3$) unified model of the DE and DR is a candidate to interpret the accelerating universe. It produces the good behaviors of cosmological quantities  and the good fits to the current observational data: growth factor and Hubble parameter. In addition, it provides an origin of the DE and DR. The energy densities of these two dark components  are self-consistent; (3)  the RMCG ($A<0$) fluid as DE also has some attractive features. For example, the CC, quintessence and phontom DE  can be realized in the RMCG ($A<0$) fluid, and in some situations the evolutions  of cosmological quantities  are not much sensitive to the variation  of model-parameters values.

  At last, we investigate the parameter space of the RMCG ($A=1/3$) and RMCG ($A<0$) model. Fitting the cosmic observational data to the  RMCG ($A=1/3$) model,  we obtain the limit on  RMCG ($A=1/3$)  model parameter $A_{s}=0.9993^{+0.0016+0.0028}_{-0.0016-0.0028}$  at 68\% and 95\% confidence levels,  which are consistent with other constraint result on $\triangle N_{eff}\in (0,1)$.  Meanwhile, the  RMCG ($A=1/3$) model almost has the same support as the most popular  $\Lambda$CDM model via the  AIC calculation. In case of fitting the cosmic data to the RMCG ($A<0$) model,  model parameters $A$ and $A_{s}$ are not convergent. The theoretical predictions on the RMCG ($A<0$) model parameters are $0< A_{s}<1$ with $-1<A<-1/3$ for the quintessence DE, and  $ 0<A_{s}<1$ with $A<-1$ for the phantom DE. But by the analysis of AIC, the RMCG ($A<0$) model has the less support from the  observational data.

 \textbf{\ Acknowledgments}
 Authors thank the Dr. Yuting Wang for improving the English of this paper. The research work is supported by the National Natural Science Foundation of China (11205078,11275035,11175077).

\section{$\text{Appendix}$}
In the following  we introduce  the cosmic data  used in this paper, including the BAO, $f_{gas}$, SNIa and H(z) data. Theoretically, one can define three distance parameter. $D_A(z)$ is the proper  angular diameter distance
\begin{eqnarray}
&&D_A(z)=\frac{c}{(1+z)\sqrt{|\Omega_k|}}\mathrm{sinn}[\sqrt{|\Omega_k|}\int_0^z\frac{dz'}{H(z')}],
\end{eqnarray}
 which relates to other two distance quantities $D_{L}$ and $D_{V}$ by
\begin{equation}
D_{L}(z)=\frac{H_{0}}{c}(1+z)^{2}D_A(z)
\end{equation}
\begin{equation}
 D_V(z)=[(1+z)^2 D^{2}_{A}(z) \frac{cz}{H(z;p_{s})}]^{1/3}
             =H_{0}[\frac{z}{E(z;p_{s})}(\int ^{z}_{0}\frac{dz^{'}}{E(z^{'};p_{s})})^{2}]^{\frac{1}{3}}.\label{eq:DV}
\end{equation}
Here $p_{s}$ is the  theoretical  model parameters, $sinn(\sqrt{|\Omega_k|}x)$ denotes $\sin(\sqrt{|\Omega_k|}x)$, $\sqrt{|\Omega_k|}x$ and
$\sinh(\sqrt{|\Omega_k|}x)$ for $\Omega_k<0$, $\Omega_k=0$ and $\Omega_k>0$,  respectively.

\subsection{$\text{BAO}$}
BAO data can be extracted from the WiggleZ Dark Energy Survey (WDWS) \cite{BAO-WiggleZ}, the Two Degree Field Galaxy Redshift Survey (2dFGRS) \cite{BAO-2dFGRs} and the Sloan Digitial Sky Survey (SDSS) \cite{BAO-SDSS}.  One can construct
\begin{equation}
\chi^2_{BAO}(p_s)=X^tV^{-1}X,\label{chi2-BAO}
\end{equation}
with
 \begin{eqnarray}
&&V^{-1}= \left(
\begin{array}{cccccc}
 4444 & 0 & 0 & 0 & 0 & 0 \\
 0 & 30318 & -17312 & 0 & 0 & 0 \\
 0 & -17312 & 87046 & 0 & 0 & 0 \\
 0 & 0 & 0 & 23857  & -22747 & 10586 \\
 0 & 0 & 0 & -22747 & 128729 & -59907 \\
 0 & 0 & 0 & 10586 & -59907 & 125536
\end{array}
\right),
X= \left(
\begin{array}{c}
 \frac{r_s(z_d)}{D_V(0.106)}-0.336 \\
 \frac{r_s(z_d)}{D_V(0.2)}-0.1905 \\
 \frac{r_s(z_d)}{D_V(0.35)}-0.1097 \\
  \frac{r_s(z_d)}{D_V(0.44)}-0.0916 \\
   \frac{r_s(z_d)}{D_V(0.6)}-0.0726 \\
    \frac{r_s(z_d)}{D_V(0.73)}-0.0592
\end{array}
\right).
\end{eqnarray}
$V^{-1}$ is the inverse covariance matrix \cite{BAO-SDSS,BAO-comatrix}. $X$ is a column vector which is given by theoretical values minus observational values, and $X^t$ denotes its transpose. $r_s(z)=c\int_0^t\frac{c_sdt}{a}=\frac{c}{\sqrt{3}}\int_{0}^{1/(1+z)}\frac{da}{a^2H(a)\sqrt{1+3a\Omega_{0b}/(4\Omega_{\gamma})}}$ is the comoving sound horizon size. $c_s^{-2}=3+\frac{4}{3}\times(\frac{\Omega_{0b}}{\Omega_{\gamma})})a$ is the sound speed of the photon$-$baryon fluid   with $\Omega_\gamma=2.469\times10^{-5}h^{-2}$.  $z_d$ denotes the drag epoch (where baryons were released from photons), $z_d=\frac{1291(\Omega_{0m}h^2)^{-0.419}}{1+0.659(\Omega_{0m}h^2)^{0.828}}[1+b_1(\Omega_{0b}h^2)^{b_2}]$ with
$b_1=0.313(\Omega_{0m}h^2)^{-0.419}[1+0.607(\Omega_{0m}h^2)^{0.674}]$ and $b_2=0.238(\Omega_{0m}h^2)^{0.223}$. $h$ is a re-normalized quantity defined by the Hubble constant $H_0 =100 h~{\rm km ~s}^{-1} {\rm Mpc}^{-1}$.

\subsection{$\text{X-ray gas mass fraction}$}
In observation of  the X-ray gas mass fraction, one can define a parameter   \cite{MCMC-fgas-data},
\begin{eqnarray}
&&f_{gas}^{\Lambda CDM}(z)=\frac{K A \gamma
b(z)}{1+s(z)}\left(\frac{\Omega_{0b}}{\Omega_{0m}}\right)
\left[\frac{D_A^{\Lambda CDM}(z)}{D_A(z)}\right]^{1.5}\ \ \ \
\label{eq:fLCDM}
\end{eqnarray}
for the reference model $\Lambda$CDM. Here $A=\left(\frac{H(z)D_A(z)}{[H(z)D_A(z)]^{\Lambda CDM}}\right)^\eta$ is the angular correction factor. $\eta=0.214\pm0.022$ is the slope of the $f_{gas}(r/r_{2500})$ data  \cite{MCMC-fgas-data}. Parameter $\gamma$ denotes permissible departures from the assumption of hydrostatic equilibrium, due to non-thermal pressure support.  Bias factor $b(z)= b_0(1+\alpha_b z)$ accounts for uncertainties in the cluster
depletion factor. $s(z)=s_0(1 +\alpha_s z)$ accounts for uncertainties of the baryonic mass fraction in stars, and a Gaussian prior for $s_0$ is employed  with $s_0=(0.16\pm0.05)h_{70}^{0.5}$ \cite{MCMC-fgas-data}. Factor $K$ is utilized to describe the combining effects of the residual uncertainties, and a Gaussian prior $K=1.0\pm0.1$ is used \cite{MCMC-fgas-data}. Adopting the datapoints  published  in Ref. \cite{MCMC-fgas-data} and following the method introduced in Refs. \cite{MCMC-fgas-data}, we can constrain theoretical model by calculating
\begin{eqnarray}
&&\chi^2_{f_{gas}}=\sum_{i=1}^{42}\frac{[f_{gas}^{\Lambda
CDM}(z_i)-f_{gas}(z_i)]^2}{\sigma_{f_{gas}}^2(z_i)}+\frac{(s_{0}-0.16)^{2}}{0.0016^{2}}
+\frac{(K-1.0)^{2}}{0.01^{2}}+\frac{(\eta-0.214)^{2}}{0.022^{2}},\label{eq:chi2fgas}
\end{eqnarray}
where $\sigma_{f_{gas}}(z_{i})$ is the statistical uncertainties. As pointed out in \cite{MCMC-fgas-data}, the acquiescent systematic uncertainties have been considered via the parameters  $\eta$, $b(z)$, $s(z)$ and $K$.

\subsection{$\text{SNIa}$}
Cosmic constraint  from SNIa observation can be determined by calculating \cite{chi2-SNIa,chi2-SNIa1,chi2-SNIa2,chi2-SNIa3,chi2-SNIa4,chi2-SNIa5,chi2-SNIa6,chi2-SNIa7,chi2-SNIa8,chi2-SNIa9,chi2-SNIa10}
\begin{eqnarray}
\chi^2_{SNIa}(p_{s})\equiv \sum_{i=1}^{557}\frac{\left\{
\mu_{th}(p_{s},z_i)-\mu_{obs}(z_i)\right\}^2} {\sigma_{\mu_{i}}^2}.\label{eq:chi2SN}
\end{eqnarray}
 Here $\mu_{obs}(z_i)$ is the observational distance moduli which can be given by  SNIa  observation datasets \cite{557Union2}, $\mu_{th}(z)=5\log_{10}[D_{L}(z)]+\mu_{0}$ is the theoretical distance modulus with $\mu_{0}=5log_{10}(\frac{H_{0}^{-1}}{Mpc})+25=42.38-5log_{10}h$, and $D_{L}(z)$ denotes the Hubble-free luminosity distance.

\subsection{$\text{ H(z) data}$}
Using the H(z) data listed in table \ref{table-29Hubbledata}, we can determine the model parameters by minimizing
 \cite{chi2hub,chi2hub1,chi2hub2,chi2hub3,chi2hub4,chi2hub5,chi2hub6,chi2hub7}
 \begin{equation}
 \chi_{H}^2(H_{0},p_{s})=\sum_{i=1}^{29} \frac{[H_{th}(H_{0},p_{s};z_i)-H_{obs}(z_i)]^2}{\sigma_{H}^2(z_i)},\label{chi2OHD}
 \end{equation}
 where  $H_{th}$ is the theoretical value and $H_{obs}$ is the observational value  for the Hubble parameter.

\end{document}